# Broadening the scope of weak quantum measurements *I*: A single particle accurately measured yet left superposed

Yakir Aharonov[1], Eliahu Cohen[1], Avshalom C. Elitzur[2]

*Weak measurement is unique in enabling measurements of non-commuting operators as well as otherwise-undetectable peculiar phenomena predicted by the Two-State-Vector-Formalism (TSVF). This article, the first in two parts, explores novel applications of weak measurement. We first revisit the basic principles of quantum measurement with the aid of the Michelson interferometer. Weak measurement is then introduced in a simple visualized manner by a specific choice of the reflecting mirror's position and momentum uncertainties. Having introduced the method, we proceed to its refinement for a single particle. We consider a photon going back and forth inside the interferometer, oscillating between a superposed and a localized state, while subjected to alternating strong and weak measurements. This cyclic process enables directly measuring both the photon's position ("which-path") and momentum (interference), without disturbing either. An alternative explanation of this result, not invoking weak values, is thoroughly considered and shown to be at odds with the experimental data. Finally a practical application of this experiment is demonstrated, where a single photon measures the various transmission coefficients of a multiport beam-splitter yet remains superposed. This method is then generalized to measurement of the wave-function itself, performed again on a single particle.*

*Part II proceeds to exploring and critically analyzing temporal peculiarities revealed by weak measurement.*

**Key Words**: weak measurement, TSVF, MZI, which-path.

# INTRODUCTION

Superposition is quantum mechanics' most intrinsic concept, an emblem of its uniqueness. An unmeasured particle's state is not only unknown but *indeterminate*, co-sustaining mutually-exclusive states. Equally crucial is the way superposition is terminated by "measurement" or "collapse," turning one of the states into an actual one, while inflicting uncertainty on conjugate variables. In view of these limitations, can there be any reason to make quantum measurement *less* precise?

[1] *School of Physics and Astronomy, Tel Aviv University, Tel Aviv 69978, Israel*

[2] *Iyar, The Israeli Institute for Advanced Research, Rehovot, Israel*

It is weak measurement [1-6], however, which can overcome these limitations, as well as many others. Moreover, the Two-State-Vector-Formalism (TSVF), within which weak measurement was conceived, predicts several very odd states occurring between ordinary measurements, which only weak measurement can reveal. Consider, for example, the very question "What is a particle's state between two measurements?" Obviously, measuring such a state would turn it into a state *upon* measurement, apparently rendering the question meaningless. Not so with weak measurement: The state can be made known with great accuracy, moreover turning out to be very unique [6].

Yet these claims are not always accepted. Critics [*e.g.* 7], while acknowledging TSVF's novelty and interest, urge it to restate its unusual claims in a simpler manner compatible with standard quantum theory.

In response to these challenges, this article's aim is twofold. It first introduces the basics of weak measurement and then broadens its scope so as to make it applicable even for a single particle. These aims constitute the present Part *I* of this paper. In Part *II* we proceed to demonstrate weak measurement's unique sensitivity to both past and future effects, again broadened to the single-particle case [8]. Both articles lay the foundations for a consecutive series of papers that explore more and more exotic features of weak measurement and TSVF [9-11].

This Part's outline is as follows. Sec. 1 revisits ordinary quantum measurement with the aid of a simple idealized setting. Within this setting, Secs. 2-3 introduce weak measurement and Secs 4-5 demonstrate its soundness for measuring otherwise-immeasurable states. Sec. 6 gives due consideration to non-TSVF accounts by which the predicted results are explained away as artifacts, and then rules them out on empirical grounds. Sec. 7 presents a refinement of weak measurement that enables it to work even with a single particle. Sec. 8 goes further to present a delicate quantum state that evades not only classical and quantum measurement, but also weak measurement as used so far, whereas the proposed refinement enables capturing that elusive state. Sec. 9 generalizes the result to the measurement of the wave function itself. Sec. 10 is a summary as well as introduction to Part *II*.

# 1. INTERFEROMETRY AND THE POSITION-MOMENTUM TRADEOFF

The Mach-Zehnder Interferometer (MZI) offers simple demonstrations of quantum superposition, quantum measurement and the uncertainty relations (Fig. 1a). The photon's initial momentum upon leaving the source, also known as the "preparation" or "pre-selection," sets the stage for the choice of the final measurement or "post-selection." Following this choice of initial momentum, a "which path" measurement constitutes a position measurement whereas interference amounts to re-measuring the photon's momentum, related by

(1) $$\Delta x \Delta p \geq \frac{\hbar}{2}.$$

Despite the MZI's simplicity, the position-momentum uncertainties tradeoff involved with its operation, as shown by Ben-Aryeh *et al.* [12], is not trivial. The following discussion rests on their analysis too.

First, the MZI setting can be further simplified by returning to its ancestor, the Michelson interferometer [13]. Let the two solid mirrors send the two split rays *back to their beam-splitter* (Fig. 1b). As long as no measurement is made to find out which path the photon took after hitting the beam splitter (BS), then (provided that the difference in the half-rays' paths is an odd multiplication of half their wavelength) the photon always returns through the BS back to the source. This is essentially an interference effect: The two returning half-rays split again at the BS, such that the two quarters going to the right cancel out, while the two going back to the source add up to the original ray's intensity.

Conversely, a "which path" measurement, even of the interaction-free type [14] which does not absorb the photon, takes its toll: It indicates whether the photon was transmitted (to the right) or reflected (to the left), but now, upon returning to the BS, it has a 50% probability to escape through the *right*-lower path (Fig. 1d). Interference and "which-path" are thus related by the same position-momentum uncertainty as (1).

Similarly for other pairs of noncommuting variables. In spin measurements, for example, SG magnets serve as beam-splitters, while interference (demonstrated by reuniting the wave function with the aid of a second SG) indicates that no which-path measurement has disturbed the spin prepared prior to the splitting.

It is this uncertainty tradeoff, apparently insurmountable, that weak measurement challenges.

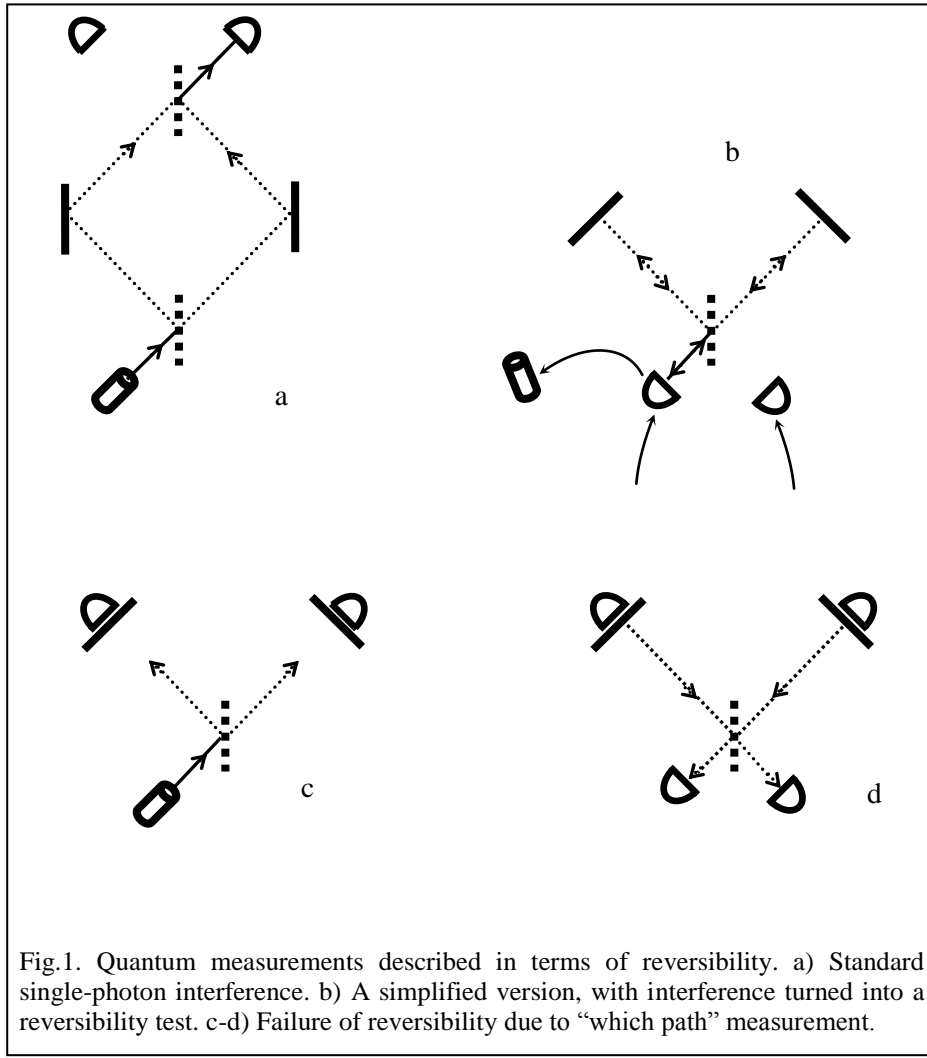

Fig.1. Quantum measurements described in terms of reversibility. a) Standard single-photon interference. b) A simplified version, with interference turned into a reversibility test. c-d) Failure of reversibility due to "which path" measurement.

## 2. MEASUREMENT: ORDINARY AND WEAK

Based on the above setting, quantum measurement can be introduced at the most fundamental level. Its weak version would then follow with equal simplicity.

The simplest measurement can be carried out by the BS and the two reflecting mirrors. The BS performs reversible *amplification* of the spatial separation between the ray's two halves ("preparation"). The reflecting mirrors, when attached to irreversible devices, complete the measuring process: Only one of them receives momentum from the photon, thereby *recording* (detecting) its position at that instant.

How, then, can this measurement be made weak? This requires compromising either the amplification or the recording stage. Most works take the former option, *i.e.*, allowing an incomplete separation between the wave function's halves. (A new method, using modulated noise, has been recently used [15] and analyzed [16].

For our purpose of measuring interference, we prefer modifying the recording stage. We therefore weaken the two reflecting mirrors' interactions with the photon. With these mirrors attached to appropriate detectors, the following holds:

**1. No Measurement:** When the mirror has a well-defined position, *e.g.*, being heavy, its momentum must be proportionately imprecise. No photon-mirror momentum transfer can then be measured, neither "which-path" information about the photon is gained. The mirrors' operation then merely time-reverses the BS's operation, leaving interference intact.

**2. Measurement:** Conversely, the mirror can be light, such that its momentum is nearly 0 with high accuracy. Any photon reflected by it imparts to it some of its momentum, disclosing its position in the right/left path. In return for its momentum precision, the mirror's position becomes uncertain, thereby ruining interference (Fig. 1d).

**3. Weak measurement:** Finally, the mirror may have some *intermediate* mass: Large enough to blur photon-mirror momentum measurement, yet sufficiently small to allow measuring some momentum transfer.

### 3. THE FORMALISM

Complementing the above introduction, let measurement (strong and weak) be put in formal terms.

First, , a few introductory comments on relevant probability laws are in order. Let a group of $N$ objects have one property in common, $s$. Even if $s$ is measured very inaccurately, the accuracy can be increased. Let each individual $s$ (the signal) be obscured by $s'$, $s''$,… (the noise), but let the overall result be summed up over all $N$ outcomes. By the Large Numbers

Law, if $x_i$ are independent and identically distributed random variables with a finite second moment, their average

$$\bar{x}_n = \frac{1}{n}(x_1 + ... + x_n),$$
(2)

goes to their expectation value $\mu$,

$$\bar{x}_n \xrightarrow{a.s} \mu.$$
(3)

As the variance is proportional to $N$, the relative error diminishes:

$$\frac{\sum_{i=1}^{N}(\Delta s)_i}{\sum_{i=1}^{N} s_i} = 1/\sqrt{N} \xrightarrow[N \to \infty]{} 0,$$
(4)

while the common factor emerges with growing precision.

Using von Neumann's arguments as in [5], a quantum measurement of the observable $A$ is defined by the interaction:

(5) $\quad H_{int}(t) = \varepsilon g(t) A P_d,$

where $P_d$ is canonically conjugated to $Q_d$, representing the pointer's position on the measuring device. The coupling $g(t)$ differs from zero only at $0 \leq t \leq T$ and normalized according to

(6) $\quad \int_0^T g(t)dt = 1,$

*i.e.* the measurement lasts no longer than $T$. Since we are interested in performing a weak measurement, we must use a small coupling strength: $\varepsilon \ll 1$.

The measurement device should be prepared in a symmetric quantum state with standard deviation $\sigma \gg \varepsilon$ and zero expectation. Without loss of generality we can refer to state $|\Psi\rangle$ in the spatial representation (which serves as our measuring base) described by a Gaussian function

$$\text{(7)} \quad \Psi(x) = \exp(-x^2/2\sigma^2).$$

The pointer movement in that case was shown [5] to be connected to the weak value of the operator $A$ defined by:

$$\text{(8)} \quad A_W = \frac{\langle \varphi | A | \psi \rangle}{\langle \varphi | \psi \rangle},$$

where $|\psi\rangle$ is the initial (preparation) state of the measured system, and $\langle \varphi |$ is the final state into which the system was projected. Hence, while weak measurement can be performed without the final state, its meaning comes only with the completion of the process, namely the combination of a past- and future-state-vectors creating the weak value of Eq. 8.

For a pre-/post-selected ensemble described by the two-state $\langle \varphi |\ |\psi\rangle$, the time evolution of the total system (measured system and the measuring device) is expected to be ($\hbar = 1$) [5]:

(9)
$$\langle \varphi | \exp(-i\int H_{int} dt) | \psi \rangle | \Psi \rangle \approx \langle \varphi | \psi \rangle (1 - i\frac{\lambda}{\sqrt{N}} A_W P_d) | \Psi \rangle = \langle \varphi | \psi \rangle \exp(-i\frac{\lambda}{\sqrt{N}} A_W P_d) | \Psi \rangle$$

which results in:

$$\text{(10)} \quad \exp(-i\frac{\lambda}{\sqrt{N}} A_W p) \Psi(x) = \Psi(x - \frac{\lambda}{\sqrt{N}} A_W).$$

For example, when weakly measuring the *spin-z* (described by the Pauli matrix $\sigma^z$) of an ensemble of spin-1/2 particles prepared in the "*x-up*" direction, with coupling strength $\varepsilon = \lambda/\sqrt{N}$, the time evolution is determined by

$$\text{(11)} \quad W = \exp(-i\int H_{int} dt) = \exp(i\lambda \sum_{n=1}^{N} \sigma_n^z P_d / \sqrt{N}),$$

so for a single measurement, the evolution of the spin states becomes entangled with the measuring device of Eq. 7 when $\sigma = 1$:

$$\text{(12)} \qquad |\sigma_z=+1\rangle e^{-(x-\lambda/\sqrt{N})^2/2} + |\sigma_z=-1\rangle e^{-(x+\lambda/\sqrt{N})^2/2}$$

*i.e.*, the individual state changes by only a fraction of $\approx \lambda^2/N$ upon strong measurement of the pointer, which will be only $\approx \lambda^2$ out of *N* particles of the ensemble. When $\lambda$ is small enough, this number of collapsed states is negligibly small compared to the ensemble's size. As collapse prevents interference we conclude that the interference pattern will be only negligibly interrupted.

It is this asymmetric tradeoff that enables weak measurement to overcome the uncertainty inflicted on the outcome by the detector's noise. Zero noise expectation and bounded variance entail noise accumulation proportional to $\sqrt{N}$, while the signal grows as *N*, hence the relative error diminishes with *N*. Thus, *an accurate result is achieved in spite of – in fact, thanks to – the measurement's weakness*. In other words, which-path information is achieved.

From Eq. 12 it follows that *O(N)* weak measurements amount to a strong measurement which almost never gives rise to collapse. A rigorous proof for this equivalence is given in [6].

Finally, two technical points need to be addressed:

**1) One or Two Path Measurements?** In an ordinary which-path measurement, it does not matter whether we place one detector on one of the MZI's paths or two on both. As one detector's clicking/non-clicking suffices, the other detector merely confirms this outcome. Not so with weak measurement: Placing two detectors amounts to performing *two* weak measurements. Moreover, these measurements, by their very nature, may well even disagree with one another, *e.g.*, giving two clicks or none. For this reason, and for the next technical issue, we opt for one detector.

**2) Individual or Collective Outcomes?** The required *N* weak measurement outcomes can be obtained either *i)* by the device interacting

with all particles, accumulating all their additive effects until giving the final outcome; or *ii)* by the device being calibrated anew after each single measurement, each outcome then individually recorded, to be summed up later under any desired grouping. In what follows we always employ *(ii)*. First, the former method inflicts an artifact in the form of a slight entanglement between the collectively measured particles. Method *(ii)*, in contrast, will prove vital for appropriately "slicing" the outcomes before re-summing them up, an advantage to be demonstrated below.

## 4. THE GAIN: MEASURING THE "IMMEASURABLE"

We are now ready to demonstrate the strength of weak measurement and TSVF. We shall do that using the simplest example - a qubit.

Consider a spin 1/2 particle undergoing two consecutive measurements of noncommuting variables, e.g., a photon's polarization measured along the co-planar directions α and β at $t_1$ and $t_2$ respectively. By the α-β uncertainty relations, repeating the α measurement at $t_3$ is no more bound to give the same outcome as in $t_1$. Now consider the intermediate time-interval $t_1<t<t_2$. What is the particle's state during this interval? The trap is obvious: You cannot answer the question by making a measurement, because then the state would rather be a state *upon* a measurement! Apparently, the problem boils down to a mere oxymoron like "measuring an unmeasured state" (as Peres' [17] sarcastically pointed out, "unperformed measurements have no results"). Weak measurement, however, enables turning the question into a physical one moreover offering a very surprising answer. Fig. 2 gives the temporal order of the process.

As pointed out above, a single weak measurement tells us nearly nothing about the state, but many such measurements enable the emergence of the weak value of Eq. 8.

$t_3$    $\sigma_\alpha^S = \downarrow$

$\sigma_\beta^W = \uparrow$
$\sigma_\alpha^W = \downarrow$

$t_2$    $\sigma_\beta^S = \uparrow$

$\sigma_\beta^W = \uparrow$
$\sigma_\alpha^W = \uparrow$

$t_1$    $\sigma_\alpha^S = \uparrow$

Fig. 2. A particle undergoing three strong measurements ($\sigma^S$) of non-commuting spin operators. Measuring the state between measurements is made possible by weak measurement ($\sigma^W$), moreover showing that this intermediate state equally agrees, at the ensemble level, with the outcomes of the past and future strong measurements.

An example of this experimental and theoretical twist is given next.

# 5. AN EXAMPLE: POSITION MEASUREMENT LEAVING MOMENTUM UNDISTURBED

Consider again the Michelson interferometer setup (Sec. 1), this time with the which-path measurement being *weak*, symbolized by the detector in Fig. 3 being gray (For the preference of a single weak detector see Sec. 2). Our aim is to show that weak position measurements retain their outcome despite a strong momentum measurement carried out between them.

First, to appreciate the weak measurement's credibility, let the BS transmission coefficient differ from the ordinary 50%, say, 55.6789%. We thus have *N* photons that share a unique state: Each photon's position is *superposed yet very specific and precise*, namely $|\psi\rangle = \sqrt{0.556789}|R\rangle + \sqrt{0.443211}|L\rangle$. Weak measurement, over time, will indeed give the 55.678-44.322% distribution of "right" and "left." With $N=10^6$, $\lambda=3$, for example, the resolution difference of $\lambda^2/N=10^{-5}$ goes to the third post-decimal digit.

Apparently, this result shows the *statistical* distribution of the *N* photons' locations, *i.e.*, their "collapses" to either side. But this is certainly not the case: *Upon the photons' returning to the BS, nearly all of them (100%-$\lambda^2$) continue back to their source, indicating interference*.

We can, conversely, remove the BS just after the upper mirror's weak measurement and prior to the photon's return to the BS (Fig. 3a). As the weak measurement's outcomes are recorded individually (Sec. 2), we can show that the two lower detectors indeed repeat the "which path" measurement, this time with a *strong* one. A weak-strong measurements' agreement is expected.

The delayed-choice version of the above two options (see Fig. 3) presents a unique feat: Weak measurement has obtained the photons' positions while leaving nearly all their momenta intact. Put differently, we have measured the BS's transmission coefficient without collapsing nearly any photon [6].

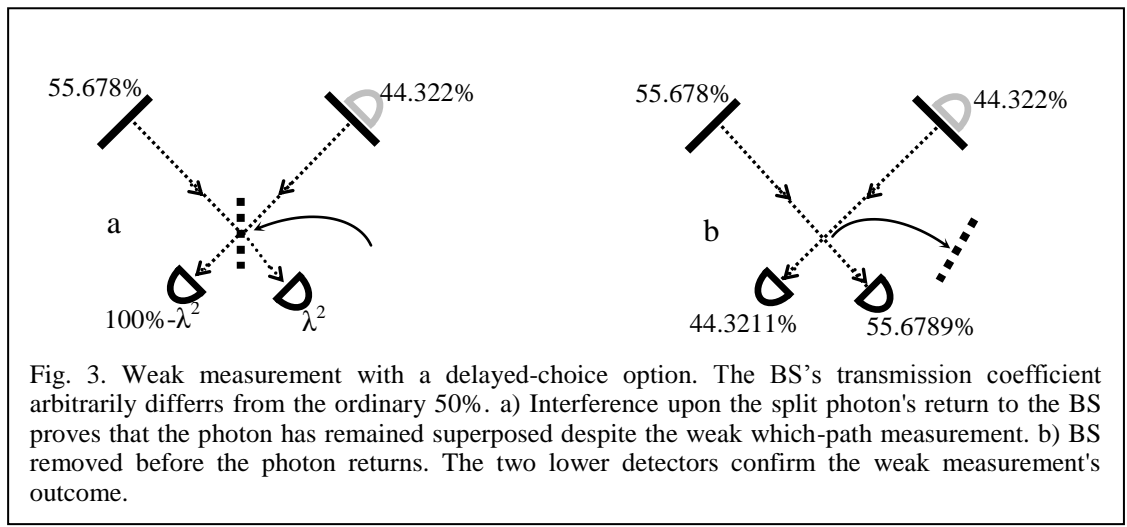

Fig. 3. Weak measurement with a delayed-choice option. The BS's transmission coefficient arbitrarily differrs from the ordinary 50%. a) Interference upon the split photon's return to the BS proves that the photon has remained superposed despite the weak which-path measurement. b) BS removed before the photon returns. The two lower detectors confirm the weak measurement's outcome.

## 6. CONSIDERING AN ALTERNATIVE ACCOUNT

As further bearings of the above results are even more peculiar [3-4], possible alternative accounts merit due hearing.

A reasonable alternative of this kind seems to be as follows. (*i*) The interference in the first choice (BS left in place) is not complete: $\lambda^2$ photons still go to the right. (*ii*) The slight deviation which gives the 55.678%-44.322% distribution in the weak measurement also differs from the real result by a mere $\lambda^2/N$. (*iii*) Perhaps, then, *it is only those $\lambda^2$ collapsed individual photons that are responsible for the weak measurement's success?*

The individual outcomes method (Sec. 2) enables ruling out this convenient alternative. Suppose we use only 1/10 of the outcomes. Since $\lambda$ can be small, *e.g.*, 2 or 3, it is quite likely that the *N/10* particles do not include the $\lambda^2$ collapsed ones. However, *N/10* being still very large, we expect the same resurrection of the photons in this case too.

It is not the few collapsed photons, then, which give the observed positions while all others display interference. Rather, *each photon* undergoes a minute change so as to perform the overall feat.

Moreover, should the experiment involve a single photon (see below), then, even one collapse would ruin the experiment, as the photon may escape the device in Fig 1b through the lower-right path.

## 7. CYCLIC WEAK MEASUREMENT: EVEN A SINGLE PARTICLE SUFFICES

We now address the challenge posed in the Introduction. Can we perform a weak measurement in which all the *N* states to be measured are of a single particle?

Recall that the technique's main strength is derived from the Large Numbers Law: By being performed on a sufficient multitude of particles, weak measurement enables obtaining almost full information about their quantum states without collapsing almost any of them. Does this mean that weak measurement must be carried out only on an ensemble of particles? In other words, should we measure *many particles* or perhaps *many states of one particle* will do? In what follows we show the latter to be the case.

Let the source in Fig. 1b emit a photon and then immediately be replaced by a mirror. This mirror is now set to function as an ordinary detector (Sec. 2), hence its measurement is strong. Now let the photon bounce back and forth between the lower and the upper mirror-detectors as in Fig. 3.b. Let it traverse the apparatus $N$ times with period $2T$, assuming negligible energy losses.

Denoting the weak measurement as in Eq. 5,

(13) $\quad H_{int} = g(t)P_d |R\rangle\langle R|/N$,

where $|R\rangle\langle R|$, the measured operator, projects the state on the right side, and $P_d$ is canonically conjugated to $Q_d$, representing the pointer position on the measuring device. The coupling $g(t)$ is non-zero only for $0 \leq t \leq \tau \ll T$ and normalized according to

(14) $\quad \int_0^\tau g(t)dt = 1$.

The measurement device is described by the Gaussian wave function mentioned above.

According to Ehrenfest's theorem,

(15) $\quad \langle \dot{Q}_d \rangle = \frac{1}{i\hbar}\langle [Q_d, H]\rangle + \left\langle \frac{\partial Q_d}{\partial t} \right\rangle = \frac{g(t)\langle |R\rangle\langle R|\rangle}{N}$.

Integrating for every integer $1 \leq m \leq N$ yields:

(16) $\quad Q_d = \int_{(2m-1)T-\tau/2}^{(2m-1)T+\tau/2} \frac{g(t)\langle |R\rangle\langle R|\rangle}{N} dt = \frac{1}{2N}$.

Hence, after $N$ steps we get $Q_d = \frac{1}{2}$, the expected value.

Thus, when $N \to \infty$, the single photon's effect on the mirror on every single interaction goes to zero, but the overall effect is measurable: A 1/2 unit of movement is transferred to the mirror. The photon's position is thereby individually measured, yet its superposition and resulting interference effect remain intact.

Naturally this setting requires high accuracy and minimal energy losses, otherwise the photon would be "collapsed" before the end of the measurements.

# 8. MAKING THE CHALLENGE HARDER: MEASURING AN UNEVEN DISTRIBUTION OF A SINGLE PHOTON'S WAVE-FUNCTION WITHOUT COLLAPSING IT

Next, consider a case where weak measurement has a clear practical advantage over all other measurements.

Alice has a beam-splitter that splits the beam into more than two parts, $n<<N$ where $N$ is the sufficient number of weak measurement outcomes to be summed up for weak measurement (Fig. 4). The *n* parts, summing up to 100%, have *unequal* intensities. Alice now asks Bob to measure all these varying intensities of her BS with maximal precision. Bob has only one photon, which, for sentimental reasons, he does not want to waste, neither even ruin its superposition. Can he do that? Four measurement methods come into account:

1. *Classical*: Out of question. You can measure the BS's transmission coefficients only with a macroscopic light beam of known initial intensity, to be compared with those of the outgoing beams. All photons, in order to be counted, must be absorbed or impart some of their energy to the detector, losing their superposition.

2. *Quantum-Mechanical*: Impossible again. You can use your photons one by one and count the number of those detected in each

outgoing beam, but even if you use mirror-detectors and the photons are not absorbed, their initial superposition is lost.

3. *Weak Measurement*: Almost there. You can use photons such that they will remain superposed even after the measurement, but they must be *as many as possible*.

4. *Cyclic Weak Measurement*: Yes you can! For your precious single photon, use the setting in Sec.3, *i.e.*, make the process cyclic: Place weak mirror-detectors on the outgoing paths. Then release your photon from its source and immediately replace the source with a mirror-detector. Now let your photon bounce back and forth within the system. This way, with strong and weak measurements cyclically alternating, you may find all Alice's beam-splitter's transmission coefficients with any desired accuracy.

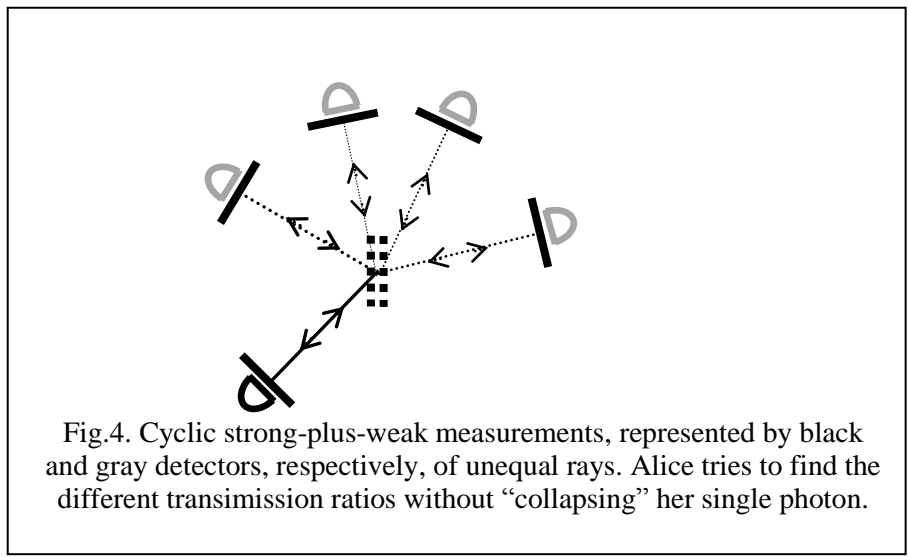
Fig.4. Cyclic strong-plus-weak measurements, represented by black and gray detectors, respectively, of unequal rays. Alice tries to find the different transimission ratios without "collapsing" her single photon.

## 9. MEASURING A SINGLE PARTICLE'S ENTIRE WAVE-FUNCTION

Making this measurement approximately continuous, *i.e.*, performed on a wave-function spread over space, amounts to *measuring the entire wave-function*. A similar result, involving protective measurement of a single particle, has been presented by Aharonov and Vaidman [18]. To apply it to our case, let a particle move through the lowest energy level of an infinite potential well located at $0 \leq x \leq L$. The probability density is known to be

(17) $|\psi(x)|^2 = \dfrac{2}{L}\sin^2(\pi x/L)$,

inside the well, and zero anywhere else. Now place two strong position detectors near $x=0$ and $x=L$, where they are expected to detect the particle very rarely (the wave function equals there almost to zero) – These are its pre- and post-selected states. In addition, place 99 weak position detectors, like the ones used above, along all intermediate locations, $x=\dfrac{L}{100}, x=\dfrac{2L}{100},..., x=\dfrac{99L}{100}$. For simplicity let $L=1$. In order to prevent wave function collapse choose a very weak coupling, say, $N=10^8$ and $\lambda=2$.

Giving the particle enough time to move back and forth in the well and very frequently collecting the detections, eventually 1000 readings are needed from each detector. Averaging the results of each detector, we find a good approximation of the position probability density (Fig. 5) with only a small probably of collapsing the wave function.

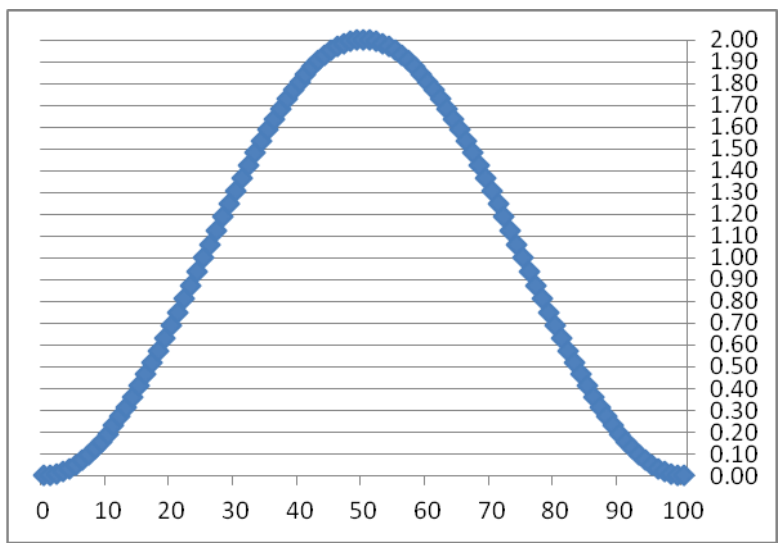

Fig.5. Simulatin of the particle probability density obtained from 1000 weak position measurements.

which is a very good estimate for the squared modulus of the wave function calculated theoretically in Eq. 17.

## 10. SUMMARY

Weak measurement involves highly sophisticated and delicate technical issues. At the conceptual level, however, it should be first studied with highly simplified settings, even *gedanken*, in order to fully appreciate its rigor and applicability.

In this article we have shown that, on such an idealized level, weak measurement can very accurately indicate a particle's state without paying the price normally exacted by the uncertainty principle. Furthermore, this method can be broadened so as to hold even for a single particle, thereby enabling measurements of states considered to be too delicate even for weak measurement as used so far.

With this refinement, other peculiar phenomena predicted by the TSVF can be demonstrated with equal simplicity, such as the past-and-future state-vector, which is the aim of Part *II*. More far-reaching applications are presented in consecutive works [8-11].

**Acknowledgements**

It is a pleasure to thank Paz Beniamini, Shay Ben-Moshe, Mannie Dorfan, Einav Friedman, Robert Griffiths, Doron Grossman and Marius Usher for helpful comments and discussions.
This work has been supported in part by the Israel Science Foundation Grant No. 1125/10.